# Discrete Rotational Symmetry, Moment Isotropy, and Higher Order Lattice Boltzmann Models

Hudong Chen[1], Isaac Goldhirsch[1,2] and Steven A. Orszag[1,3]

## Abstract

Conventional lattice Boltzmann models only satisfy moment isotropy up to fourth order. In order to accurately describe important physical effects beyond the isothermal Navier-Stokes fluid regime, higher-order isotropy is required. In this paper, we present some basic results on moment isotropy and its relationship to the rotational symmetry of a generating discrete vector set. The analysis provides a geometric understanding for popular lattice Boltzmann models, while offering a systematic procedure to construct higher-order models.

[Key words: lattice Boltzmann method; moment isotropy; discrete rotational symmetry; fluid dynamics]

[1] Exa Corporation, 3 Burlington Woods Drive, Burlington, MA 01821
[2] Also, School of Mechanical Engineering, Tel-Aviv University, Ramat-Aviv, ISRAEL
[3] Also, Department of Mathematics, Yale University, New Haven, CT 06510

# 1. Introduction

The Lattice Boltzmann Method (LBM) is increasingly recognized as an effective method for numerical fluid flow simulations [2, 12 18]. Besides its algorithmic simplicity, capability to be efficiently parallelizable, high numerical stability [7, 23], and effectiveness for time-dependent fluid computations, the representation of fluid flows by the LBM has a physical basis that results from its kinetic theoretical origins. Indeed, the LBM is arguably applicable to a wider variety of fluid flow phenomena than conventional macroscopic (hydrodynamic) based descriptions [1, 5, 17], as LBM may capture phenomena on mesoscopic scales.

There have been extensive studies of LBM over the past last decade. However, one important aspect of LBM has not yet received sufficient notice: Because a handful of discrete velocity values are employed in LBM, the advection process in LBM can be explicitly carried out for each and every distribution at the specific velocity values that are used. By directly advecting these distribution functions, both the thermodynamic equilibrium and the full non-equilibrium contributions (in this representation) are retained and transported without error. This is essential for flows that contain substantial non-equilibrium contributions to their distribution functions. In comparison, hydrodynamic-based approaches, such as the Navier-Stokes equation, rely on approximations for non-equilibrium effects which are not valid when deviations from local thermodynamic equilibrium are significant, i.e. when gradients of the hydrodynamic fields are "large" or their time dependence is "fast".

The LBM is in principle not limited to regimes involving only small deviations from equilibrium. On the other hand, although all important non-equilibrium effects are contained in the continuous Boltzmann equation (for dilute gases) and the Enskog-Boltzmann equation (for moderately dense gases), it is virtually impossible numerically to perform the advection of distributions for all continuous velocities. For that reason, indirect means such as moment flux-based schemes are sought and these encounter the problem of not being easily represented by approximate analytical expressions (cf. [29, 30]).

A gratifying feature of LBM is that, due to the precise synchronization between the underlying lattice spacing, the set of constant velocity values and the time step, the advection process in standard LBM is exact. There is no numerical diffusion generated by the advection process. With the new multi-speed schemes developed herein, it is important to note there is no need to introduce fractional time-stepping because of the nature of the induced spatial grids (see below and also [31]).

Although the theoretical range of validity for the LBM may be rather broad, most existing LBM models are only accurate within the Navier-Stokes regime. While it can be claimed that the LBM captures non-equilibrium effects beyond the Navier-Stokes regime, the corrections to the Navier-Stokes description are often contaminated by numerical artifacts caused by the finite and discrete velocity space. In fact, the discovery that fourth-order moment isotropy of a discrete velocity set leads to the isothermal Navier-Stokes equations was a celebrated turning point in the development of lattice gas models [16, 28]. The fact that simulations of Navier-Stokes flows *via* the LBM have been the overwhelming application of this method up to now has led to some confusion concerning the domain of applicability of LBM. It is therefore important to formulate theoretical extensions to LBM models that can cover a broader fluid flow regime beyond the Navier-Stokes regime [22]. One of the central tasks is to define and construct extended discrete velocity sets that have higher-order moment isotropy. In this paper, we present a theoretical study that is directly relevant to the achievement of this goal.

Specifically we focus here on the rotational symmetries for discrete lattice velocity vector sets and their relationships to the orders of moment isotropy. Furthermore, we present a theoretical procedure to evaluate and systematically construct extended isotropic discrete velocity sets that can then be used for higher-order LBM models. The concept of rotational symmetry and moment isotropy was extensively discussed earlier by Wolfram [28], a cornerstone of lattice gas models and the majority of the LBM models for Navier-Stokes fluid flows. On the other hand, the fundamental (and important) issue of going beyond the Navier-Stokes regime, has not been thoroughly explored. The analysis presented here also provides some further explanation of the geometric origins of some popularly known lattice Boltzmann models.

## 2. Rotational invariance and hydrodynamics

A lattice Boltzmann equation (LBE) is commonly expressed in finite-difference form as follows [2, 13]

$$f_\alpha(\mathbf{x}+\mathbf{c}_\alpha, t+1) - f_\alpha(\mathbf{x},t) = (\partial f / \partial t)_c \qquad (1)$$

where $f_\alpha(\mathbf{x},t)$ represents the number density of particles with velocity value $\mathbf{v} = \mathbf{c}_\alpha$ at $(\mathbf{x},t)$. In the Bhatnagar-Gross-Krook (BGK) approximation, the collision operator $(\partial f / \partial t)_c$ is [3,9,19]

$$(\partial f / \partial t)_c = -\frac{1}{\tau}[f_\alpha(\mathbf{x},t) - f_\alpha^{eq}(\mathbf{x},t)] \qquad (2)$$

where $1/\tau$ models the collision rate. Without loss of generality, one adopts "lattice units" in which the basic time increment (time step) is unity, $\Delta t = 1$. There are typically only $b$ ($\sim 20$) fixed velocity values in an LBE system that form a discrete lattice vector set $C$ ($\equiv \{\mathbf{c}_\alpha; \alpha = 1,...,b\}$). Each $\mathbf{c}_\alpha$ is a vector in $D$-dimensional space, (typically, $D = 2$, $3$, and $4$) and serves as an exact link between two neighboring sites on a Bravais lattice. Each LBM model has a corresponding specified lattice velocity set, determined by the time step.

Hydrodynamic moments in LBM are represented by summations over the discrete lattice velocity values. The most familiar moments are the ones corresponding to the mass and momentum of the fluid, respectively

$$\rho(\mathbf{x},t) \equiv \sum_{\alpha=1}^{b} f_\alpha(\mathbf{x},t), \quad \rho\mathbf{u}(\mathbf{x},t) \equiv \sum_{\alpha=1}^{b} \mathbf{c}_\alpha f_\alpha(\mathbf{x},t) \qquad (3)$$

where $\rho$ and $\mathbf{u}$ are, respectively, the local fluid density and velocity. Similarly, the (kinetic) momentum flux tensor $P_{ij}$ and the energy flux tensor $Q_{ijk}$ can also be defined,

$$P_{ij}(\mathbf{x},t) \equiv \sum_{\alpha=1}^{b} c_{\alpha,i} c_{\alpha,j} f_\alpha(\mathbf{x},t) \qquad (4)$$

$$Q_{ijk}(\mathbf{x},t) \equiv \sum_{\alpha=1}^{b} c_{\alpha,i} c_{\alpha,j} c_{\alpha,k} f_\alpha(\mathbf{x},t) \qquad (5)$$

where sub-indices $i$, $j$ and $k$ denote Cartesian components of a vector (or tensor) in $D$-dimensional space. Higher order moments can be similarly defined.

From the above, we see that properties of hydrodynamic moments depend on the choice of the discrete vector set $C$. In fact, all hydrodynamic moments are ultimately expressible in terms of the basis moment tensors defined below: An $n^{th}$ order basis moment tensor has the generic form,

$$\mathbf{M}^{(n)} = \sum_{\alpha=1}^{b} w_\alpha \underbrace{\mathbf{c}_\alpha \otimes \mathbf{c}_\alpha \otimes \cdots \otimes \mathbf{c}_\alpha}_{n} \qquad (6)$$

where the set of weights $\{w_\alpha\}$ can be different for velocities of different magnitude (speed) and $\otimes$ indicates tensor product. Note that even though the distribution function $f_\alpha$ can be weakly $(\mathbf{x},t)$-dependent, the weights $\{w_\alpha\}$ can still be chosen independent of $(\mathbf{x},t)$. In cases involving rapid change in $(\mathbf{x},t)$ [say, e.g., due to rapid change of temperature $T$ in a non-isothermal fluid], the weights $\{w_\alpha\}$ would be chosen to be functions of $T$.

In terms of Cartesian components, the form of $M^{(n)}$ for $n \geq 0$ is

$$M^{(n)}_{i_1 i_2 \ldots i_n} \equiv \sum_{\alpha=1}^{b} w_\alpha\, c_{\alpha,i_1} c_{\alpha,i_2} \cdots c_{\alpha,i_n} \qquad (7)$$

where the sub-indices $i_1, \ldots, i_n$ denote Cartesian components. In fact, $M^{(n)}_{i_1 i_2 \ldots i_n}$ is a symmetric tensor under permutation of any pair of its Cartesian indices. The transformations under rotation of all the hydrodynamic moments are fully determined by the basis moment tensors.

A discrete vector set $C$ naturally admits a geometric interpretation: It forms a crystallographic structure in $D$-dimensional space with all vectors in $C$ starting from the same origin and pointing in their own respective directions. In addition, this crystallographic structure forms a "rigid body" in $D$-dimensional space. Clearly, the intrinsic symmetry properties of such a rigid body crystallographic structure does not depend on its relative orientation with respect to the coordinate system. On the other hand, each component of a representation of the $n^{th}$ order moment tensor $\mathbf{M}^{(n)}$ is in general expected to vary as one varies the orientation of $C$. For a crystallographic structure that possesses a discrete rotational symmetry, $\mathbf{M}^{(n)}$ must be invariant under the corresponding discrete rotations [15,20,26]. Moreover, if $C$ is itself a Bravais lattice, it is parity symmetric (i.e., $\mathbf{c}_\alpha \to -\mathbf{c}_\alpha$, $\forall \mathbf{c}_\alpha \in C$). Thus, we conclude that $\mathbf{M}^{(n)}$ is invariant under parity transformation, and all odd integer moments ($n$ odd) vanish.

It is clear that the fact that a given vector set is invariant under a finite set of discrete rotations does not imply that it obeys all symmetry requirements of continuum physics, as the latter clearly demands no preference given to any particular coordinate orientations. In other words, because there is a continuum of velocities in a real physical system, the basis moment tensors (7) of all orders in such a system are invariant under arbitrary (proper and improper) rotations rather than just a few particular discrete rotations. As a consequence, $\mathbf{M}^{(n)}$ for any even integer $n$ must be an isotropic tensor. In other words, the result of a repeated scalar product $\mathbf{M}^{(n)}$ with a $D$-dimensional vector $\mathbf{V}$ of arbitrary orientation must be a function of only the magnitude of $\mathbf{V}$:

$$\mathbf{M}^{(n)} \cdot \mathbf{V} = \sum_{\alpha=1}^{b} w_\alpha \sum_{i_1=1}^{D} \sum_{i_2=1}^{D} \cdots \sum_{i_n=1}^{D} c_{\alpha,i_1} c_{\alpha,i_2} \cdots c_{\alpha,i_n} V_{i_1} V_{i_2} \cdots V_{i_n}$$

$$= \sum_{\alpha=1}^{b} w_\alpha (\mathbf{c}_\alpha \cdot \mathbf{V})^n \propto V^n \qquad (8)$$

where $V \equiv |\mathbf{V}| = \sqrt{V_1^2 + V_2^2 + \cdots + V_D^2}$ is the magnitude of the vector. It is known that up to a trivial scalar multiplicative factor, a parity invariant $n^{th}$ order isotropic tensor has the form of the $n^{th}$ order 'isotropic delta function' $\Delta^{(n)}_{i_1 i_2 \ldots i_n}$ [4, 20, 26, 28]. The latter is defined as the sum of $n/2$ products of simple Kronecker delta functions $\delta_{i_1 i_2} \cdots \delta_{i_{n-1} i_n}$ over all distinctive permutations of its sub-indices. There are $(n-1)!!$ ($\equiv (n-1)\cdot(n-3)\ldots 3\cdot 1$) distinctive terms in $\Delta^{(n)}_{i_1 i_2 \ldots i_n}$. For instance, $\Delta^{(2)}_{ij} \equiv \delta_{ij}$, and

$$\Delta^{(4)}_{ijkl} = \delta_{ij}\delta_{kl} + \delta_{ik}\delta_{jl} + \delta_{il}\delta_{jk}$$

$$\Delta^{(6)}_{ijklmn} = \delta_{ij}\Delta^{(4)}_{klmn} + \delta_{ik}\Delta^{(4)}_{jlmn} + \delta_{il}\Delta^{(4)}_{jkmn} + \delta_{im}\Delta^{(4)}_{jkln} + \delta_{in}\Delta^{(4)}_{jklm} \qquad (9)$$

One cannot expect to achieve full rotational symmetry for all $n$ from a $C$ that comprises only a finite number of vectors. However, one *can* expect isotropy from $C$ up to some finite order $n \leq N$ with $N$ dependent on the velocity set $C$. For example the sets corresponding to the simple 2D square and 3D cubic lattices admit $N = 2$, while the 2D hexagonal lattice and the 4D face-centered hypercube (4D FCHC) lattice achieve $4th$ order isotropy ($N = 4$) [16, 28]. This is important because many hydrodynamic phenomena are describable by the isothermal Navier-Stokes equations at low Mach number and an LBM for such flows is known to only require a $4^{th}$ order isotropy [9, 12, 16, 19, 28]. However, $N > 4$ is expected to be required when one wishes to go beyond the isothermal Navier-Stokes regime. Indeed, based on a Chapman-Enskog-like analysis, one can show that $N \geq 6$ is necessary to realize the correct Navier-Stokes level thermohydrodynamics or Burnett level (in the framework of kinetic-theory-based hydrodynamics) isothermal fluid physics [6, 10, 21, 22].

While some $6^{th}$ order isotropic LBM models exist for non-isothermal Navier-Stokes flows, there has been no systematic study yet concerning the generation of higher order isotropic lattices [14, 25, 27].

## 3. Basic symmetry properties of parity invariant lattices

Let us begin by examining a simple class of lattices in which all velocities in $C$ have the same magnitude (speeds), $|\mathbf{c}_\alpha| = c$. It is apparent that to achieve maximum rotational symmetry, we should choose the same value for $w_\alpha$ among all velocities. Furthermore, we can choose ($w_\alpha = 1$) without loss of generality. More general lattices involving multiple speed levels can be straightforwardly constructed by combining multiples of these simple lattices, or projections of these on to lower dimensional spaces (see Appendix 2). The most commonly known examples of such simple lattices are the 2D square lattice, 2D hexagonal lattice, and the 4D-FCHC lattice [16, 28]. For these simple lattices, one can show (*cf.*, [28]) *via* straightforward summations that an isotropic $n^{th}$ order tensor $M^{(n)}_{i_1 i_2 \ldots i_n}$ has the following form:

$$M^{(n)}_{i_1 i_2 \ldots i_n} = bc^n \frac{(D-2)!!}{(D+n-2)!!} \Delta^{(n)}_{i_1 i_2 \ldots i_n}; \quad n = 2, 4, 6, \ldots \tag{10}$$

where $(D+n-2)!!/(D-2)!! \equiv (D+n-2) \cdot (D+n-4) \ldots (D+2) \cdot D$ and $b$ is the number of vectors in $C$. Derivations of this expression for certain specific orders is given in Appendix 3.

Accordingly, for $D \geq 2$ we can deduce a set of hierarchical relationships among the tensor components at different orders:

$$M^{(n)}_{\underbrace{i \cdots i}_{n}} = \frac{(n-1)c^2}{D+n-2} M^{(n-2)}_{\underbrace{i \cdots i}_{n-2}}$$

$$M^{(n)}_{\underbrace{i \cdots i}_{n-2} jj} = \frac{(n-3)c^2}{D+n-2} M^{(n-2)}_{\underbrace{i \cdots i}_{n-4} jj}$$

$$M^{(n)}_{\underbrace{i \cdots i}_{n-4} jjjj} = \frac{(n-5)c^2}{D+n-2} M^{(n-2)}_{\underbrace{i \cdots i}_{n-6} jjjj} \tag{11}$$

and so on, where the indices $i \neq j$ are any Cartesian components and the summation convention over repeated indices is *not* used here and below. Furthermore, one can obtain an additional relationship between higher and lower order moments by contracting the higher order ones,

$$\sum_j M^{(n)}_{i_1 i_2 \ldots i_{n-2} jj} = c^2 M^{(n-2)}_{i_1 i_2 \ldots i_{n-2}} \tag{12}$$

Equation (12) shows that isotropy conditions at lower orders are automatically satisfied if there is isotropy at a higher order, as could be intuitively expected.

The relationships (11) with (12) lead to a set of necessary and sufficient conditions to determine the maximal order of isotropy for any given simple lattice. A good way to do this is to determine how many such constraint relations from (11) and (12) are needed to characterize isotropy at each given order. This is done by finding additional basic properties of a lattice vector set that is parity invariant and rotational symmetric in Cartesian coordinates. For such a lattice, $M^n_{i_1 \ldots i_n}$ is zero whenever a particular component index appears an odd number of times. For example, $M^n_{i_1 \ldots i_n}$ is identically zero for any odd integer $n$, since at least one index in it will appear an odd number of times. Furthermore, one can argue as follows for a tensor of even integer rank. $M^n_{i_1 \ldots i_n}$ ($n$ even) is non-zero if and only if either: (1) All its Cartesian indices are equal; or (2) The indices have more than one distinct value, but each value appears even number of times. By analyzing all possible permutations of indices, one concludes that there are $(n-1)!!$ non-vanishing components for a tensor of even integer rank $n$. With a symmetric tensor, such as $\mathbf{M}^{(n)}$, its values must be the same under arbitrary permutation of indices. Thus, these tensors have the following generic forms (*cf.* [28]):

$$M^{(2)}_{ij} = \frac{bc^2}{D} \delta_{ij}$$

$$M^{(4)}_{ijkl} = \frac{bc^4}{D(D+2)} [\Phi^{(4)}_0 \Delta^{(4)}_{ijkl} + \Phi^{(4)}_1 \delta_{ijkl}]$$

$$M^{(6)}_{ijklmn} = \frac{bc^6}{D(D+2)(D+4)} [\Phi^{(6)}_0 \Delta^{(6)}_{ijklmn} + \Phi^{(6)}_1 \delta_{ijklmn} + \Phi^{(6)}_2 \delta^{(4,2)}_{ijklmn}]$$

$$M^{(8)}_{ijklmnpq} = \frac{bc^8}{D(D+2)(D+4)(D+6)} [\Phi^{(8)}_0 \Delta^{(8)}_{ijklmnpq} + \Phi^{(8)}_1 \delta_{ijklmnpq}$$

$$+ \Phi^{(8)}_2 \delta^{(6,2)}_{ijklmnpq} + \Phi^{(8)}_3 \delta^{(4,4)}_{ijklmnpq} + \Phi^{(8)}_4 \delta^{(2,2,4)}_{ijklmnpq}] \tag{13}$$

where the coefficients, $\{\Phi^n_\alpha\}$, are completely determined by the geometric structure of each given lattice. Besides the isotropic delta functions $\Delta^{(n)}_{i_1 \ldots i_n}$, there appear 'atypical' Kronecker delta functions of $n^{th}$ order, $\delta_{i_1 \ldots i_n}$ with $n$ subindices: here $\delta_{i_1 \ldots i_n} = 1$ if all its subindices are equal, while $\delta_{i_1 \ldots i_n} = 0$ otherwise. The quantity $\delta^{(n-m,m)}_{i_1 \ldots i_n} (m = 0, 2, \ldots, n)$ is a direct product of two Kronecker delta functions defined as the sum of the product of $\delta_{i_1 \ldots i_{n-m}} \delta_{i_{n-m+1} \ldots i_n}$ over all possible distinct permutation of sub-indices. Similarly, $\delta^{(n_1, n_2, n_3)}_{i_1 \ldots i_n}$ ($n_1 + n_2 + n_3 = n$) is a direct product of three Kronecker delta functions. According to the expressions for isotropic tensors (10), we can conclude that a lattice vector set satisfying moment isotropy up to $N^{th}$ order must not contain any of these atypical delta-functions. In order words, the coefficient conditions $\Phi^{(n)}_0 = 1$ and $\Phi^{(n)}_{k>0} = 0$ for all $n \leq N$ must be satisfied in a $N^{th}$ order isotropic lattice set.

From (13), we see that this type of lattice is always *2nd* order isotropic. Also, there are only two unknown coefficients for the *4th* order tensor, three for the *6th* order, five for the *8th* order, and so on. Therefore, we can choose the appropriate number of independent constraint relations from (11) and (12) to evaluate and determine the order of isotropy of a given lattice. In particular, from (13)

$$M^{(4)}_{iiii} = \frac{bc^4}{D(D+2)} [3\Phi^{(4)}_0 + \Phi^{(4)}_1] \tag{14}$$

where $i$ can be any Cartesian component *x, y*, or *z*. Hence, to verify if a given lattice is $4^{th}$ order isotropic, we only need to use two independent constraint relations. Furthermore, since such a lattice set is always $2^{nd}$ order isotropic, from (12) we realize that one constraint is already automatically satisfied between the two unknown coefficients,

$$\Phi^{(4)}_0 + \frac{\Phi^{(4)}_1}{D+2} = 1, \tag{15}$$

so that only one additional constraint relationship needs to be selected and checked. Using the first hierarchical relation in (11), we can immediately obtain an additional constraint, namely

$$M^{(4)}_{iiii} = 3 \frac{bc^4}{D(D+2)} \tag{16}$$

To check isotropy at $4^{th}$ order, (16) provides a sufficient condition. Directly applying it, we can verify that the 2D hexagonal and 4D-FCHC lattices are $4^{th}$ order isotropic, while the 2D square and 3D cubic lattices are not.

Relationships (11) and (12) also provide a set of systematic measures for successively determining isotropy characteristics of a lattice at higher orders. For instance, if a given lattice is already known to be $4^{th}$ order isotropic, we can continue to examine its $6^{th}$ order properties using (11)-(12). Contracting $M^{(6)}$ to 4$^{th}$ order and using (12), we arrive at the two constraints,

$$\Phi_0^{(6)} + 2\frac{\Phi_1^{(6)}}{D+4} = 1$$

$$\Phi_1^{(6)} + (D+8)\Phi_2^{(6)} = 0 \tag{17}$$

so that only one coefficient is independent in $M^{(6)}$. Therefore, isotropy at 6$^{th}$ order is determined by checking just one additional isotropy condition. This can be chosen either from the first or the second hierarchical relationship in (11), namely

$$M_{iiiiii}^{(6)} = \frac{5c^2}{D+4}M_{iiii}^{(4)} = 15\frac{bc^6}{D(D+2)(D+4)}$$

$$M_{iiiijj}^{(6)} = \frac{3c^2}{D+4}M_{iijj}^{(4)} = 3\frac{bc^6}{D(D+2)(D+4)} \tag{18}$$

where $i \ne j$ can be any Cartesian components. From (13), the constraints can also be expressed in the alternative simpler forms below,

$$M_{iiiiii}^{(6)} = \frac{bc^6}{D(D+2)(D+4)}[15\Phi_0^{(6)} + \Phi_1^{(6)} + 15\Phi_2^{(6)}]$$

$$M_{iiiijj}^{(6)} = \frac{bc^6}{D(D+2)(D+4)}[3\Phi_0^{(6)} + \Phi_2^{(6)}] \tag{19}$$

In conclusion, if a given lattice is known to be isotropic up to 4$^{th}$ order, then this lattice is also isotropic at 6$^{th}$ order provided that one of the conditions in (18) or (19) is satisfied.

Based on the above analysis, a general procedure may be described: For a lattice vector set belonging to the simple class, if it is known to be isotropic at the $n^{th}$ order, then its moment isotropy at the $(n+2)^{nd}$ order can be determined by examining one of the $(n+2)^{nd}$ order constraint relations in (11).

There is an alternative, and more straightforward procedure, to verify the order of isotropy for a given lattice. In fact, from (9) we find that the following hierarchical relations are satisfied for an $n^{th}$ order isotropic tensor:

$$M_{\underbrace{i_1 \cdots i_1}_{n}}^{(n)} = (n-1)M_{\underbrace{i_1 \cdots i_1}_{n-2}i_2 i_2}^{(n)}$$

$$= (n-1)(n-3)M_{\underbrace{i_1 \cdots i_1}_{n-4}i_2 i_2 i_3 i_3}^{(n)} = \ldots$$

$$= ((n-1)!!)M_{i_1 i_1 i_2 i_2 \cdots i_D i_D}^{(n)} \tag{20}$$

where the indices $i_1 \ne i_2 \ne \cdots \ne i_D$ can be any given Cartesian components. Specifically, for $n = 4$, we have

$$M^{(4)}_{iiii} = 3M^{(4)}_{iijj}$$

Similarly,

$$M^{(6)}_{iiiiii} = 5M^{(6)}_{iiiijj} = 15M^{(6)}_{iijjkk}$$

where $i \neq j \neq k$. Once again, since such lattices are invariant under (proper and improper) Cartesian coordinate transformations and all Cartesian components are equivalent, it is sufficient to simply choose a particular set of Cartesian components for verifying the isotropy order of a lattice. For instance, we can pick $i = x$, $j = y$, $k = z$, and so on. Once we verify the above relationships are satisfied by these particular components, all the other conditions are automatically satisfied. This alternative procedure provides a sufficient and convenient way to examine isotropy properties for a given lattice. Note that this alternative procedure is also applicable to lattice sets involving multiple speeds.

## 4. Increased order of isotropy using combinations of lattices

In this Section, we describe a systematic procedure to construct higher order lattice sets using combinations of lower order lattice sets. Generally speaking, it can be shown that an $N^{th}$ order isotropic lattice set can be generated out of a union of a $(N\text{-}2)^{nd}$ order isotropic lattice set and its rotated realizations. We demonstrate this *via* specific representative examples.

### a. Fourth order isotropy based on rotated 2D square lattices

The standard 2D square lattice is comprised of the following ($b = 4$, $D = 2$) lattice velocities (of all unity magnitude, $c = 1$),

$$C_2 = \{(\pm 1, 0), (0, \pm 1)\}, \tag{21}$$

where here and below $\{(\pm 1, 0)\}$ is a shorthand notation for $\{(+1,0), (-1,0)\}$, and $\{(\pm 1, \pm 1)\} \equiv \{(+1,+1), (+1,-1), (-1,+1), (-1,-1)\}$.

It is known that the square lattice is only $2^{nd}$ order isotropic. Indeed, this is directly verified *via* the diagonal component of its $4^{th}$ order moment tensor, $M^{(4)}_{iiii}$. Because all the Cartesian components are equivalent, with no loss of generality, the index $i$ can be chosen for convenience to be $x$ and the following is easily calculated

$$M^{(4)}_{iiii} = M^{(4)}_{xxxx} = \sum_{\alpha=1}^{b=4} c^4_{\alpha, x}$$

$$= 1^4 + 0^4 + (-1)^4 + 0^4 = 2 \tag{22}$$

Clearly this does not satisfy the basic isotropy requirement at $4th$ order according to (16) as the latter requires

$$M^{(4)}_{iiii} = 3\frac{bc^4}{D(D+2)} = \frac{3}{2}$$

using $D = 2$, $b = 4$ and $c = 1$.

The 2D square lattice set (21) is geometrically a set of four mutually orthogonal vectors in a 2-dimensional space. A crystallographically identical set can be generated by simply performing a rigid-body rotation of this standard set (21). In particular, if the angle of rotation is $\pi/4$, the resulting rotated set $C_2^+$ is comprised of the following four new velocity vectors:

$$C_2^+ \equiv \{(\pm 1/\sqrt{2}, \pm 1/\sqrt{2})\} \tag{23}$$

Being merely a rotation, obviously all tensor properties in $C_2^+$ are by definition identical to those in the original 2D square lattice set, and thus isotropy properties in $C_2^+$ are the same as those in $C_2$. However, as expected the Cartesian components for each moment tensor do change due to the rotation. A direct calculation reveals

$$M_{iiii}^{(4)} = 1 \neq \frac{3}{2} \tag{24}$$

so the rotated 2D square lattice is also not 4$^{th}$ order isotropic.

Next, we can form a super set $B_2$ as a union of the two 2D square lattice sets, $C_2$ and $C_2^+$:

$$B_2 \equiv C_2 \cup C_2^+ \tag{25}$$

in which there are eight 2D vectors ($b = 8$). Evaluating its 4$th$ order moment gives

$$M_{iiii}^{(4)} = 3 \tag{26}$$

which agrees with the sufficient isotropy condition (16):

$$3\frac{bc^4}{D(D+2)} = 3$$

Using $D = 2$, $c = 1$, and $b = 8$. Therefore, the new super set $B_2$ achieves a higher order isotropy ($N = 4$) than its two subsets each of which only satisfy 2$^{nd}$ order isotropy.

From $B_2$, we can make an interesting observation: The expanded vector set corresponds geometrically to the standard 2D "octagonal" lattice which possesses an eight-fold rotational symmetry as opposed to a four-fold symmetry in each of the original 2D square lattices. In other words, the new lattice set is invariant under rotations of angles integer multiples of $2\pi/8$. More generally, we say that a 2D lattice is $m$-degree (or $m$-fold) symmetric if the velocity set is invariant under a rigid-body rotation of angles that are integer multiples of $2\pi/m$. In fact, the above procedure for creating super sets can be further carried out to include $M$ ($\geq 2$) number of rotated square lattice sets, each of which is a realization of $\pi/2M$ rotation of the other. The resulting superset thus has an $m$ ($= 4M$)-fold rotational symmetry.

There exists a general relationship between the rotational symmetry of a 2D vector set and the moment isotropy at $n^{th}$ order:

*Theorem: If a 2D lattice set is $(m+2)$-degree symmetric, then its moments are isotropic up to $m^{th}$ order. Here $m$ ($\geq 0$) is an even integer.*

The detailed proof is given in Appendix 1. Using this general relationship, we immediately conclude that a 2D square lattice is only 2$^{nd}$ order isotropic, while the well known hexagonal lattice is 4$th$ order isotropic for it has a 6-degree ($m = 6$) rotational symmetry. Furthermore, being 8-degree symmetric ($m = 8$) the octagonal lattice above is expected to be isotropic up to 6$^{th}$ order instead of only 4$^{th}$. Directly evaluating the 6$^{th}$ order moment on $B_2$, we find that the sufficient condition (18) is indeed satisfied,

$$M^{(6)}_{iiiiii} = \frac{5}{2} = 15\frac{bc^6}{D(D+2)(D+4)} \tag{27}$$

*b. Sixth-order isotropy via a combination of 4D-FCHC*

We can further examine lattice vector sets of higher dimensionalities ($D > 2$). Unfortunately, the general relationship between symmetry and isotropy for 2D cannot be directly extended to higher dimensions, due to the finiteness of the number of "Platonic shapes" in higher dimensions (*cf.* [15]). Nevertheless, following the same procedure as described in the previous subsection, we present a way to construct a 6$^{th}$ order isotropic lattice from two crystallographically equivalent 4D-FCHC lattices. By definition, the standard 4D-FCHC lattice set is given (in terms of Cartesian components) by the following $b$ ($= 24$) velocity values in $D$ ($= 4$) dimensional space,

$$C_4 \equiv \{(\pm 1, \pm 1, 0, 0), (\pm 1, 0, \pm 1, 0), (\pm 1, 0, 0, \pm 1),$$
$$(0, \pm 1, \pm 1, 0), (0, \pm 1, 0, \pm 1), (0, 0, \pm 1, \pm 1)\} \tag{28}$$

All velocities in the set have the same magnitude $c = \sqrt{2}$. It is well known that this FCHC lattice is isotropic up to 4$^{th}$ order [16, 28], and its moments at 6$^{th}$ order and higher are not isotropic. Indeed, direct evaluating its 6$^{th}$ order moment, we find

$$M^{(6)}_{iiiiii} = 12 \neq 15 = 15\frac{bc^6}{D(D+2)(D+4)} \tag{29}$$

confirming that the basic isotropy relationship (18) is violated.

As for 2D lattices, a crystallographically identical 4D-FCHC lattice vector set can be realized *via* a particular rigid-body rotation of the original standard FCHC lattice vector set. In particular, if the rotation angle is chosen to be "$\pi/4$" in 4D, a new FCHC lattice set $C_4^+$ is generated: The 24 rotated new velocities expressed in terms of the Cartesian components are given below,

$$C_4^+ = \{(\pm\sqrt{2}, 0, 0, 0), (0, \pm\sqrt{2}, 0, 0), (0, 0, \pm\sqrt{2}, 0), (0, 0, 0, \pm\sqrt{2}),$$
$$(\pm\frac{1}{\sqrt{2}}, \pm\frac{1}{\sqrt{2}}, \pm\frac{1}{\sqrt{2}}, \pm\frac{1}{\sqrt{2}})\} \tag{30}$$

In fact, the resulting lattice is the dual lattice of the standard 4D-FCHC [15]. Being crystallographically identical and merely a rotation, this new set has all the same symmetry properties as the original 4D-FCHC. Directly computing its specific 6$^{th}$ order moment component, we have

$$M^{(6)}_{iiiiii} = 18 \neq 15 \tag{31}$$

where $i$ can be any of its four Cartesian components, so it is not isotropic at 6$^{th}$ order.

Once again, we can form a "super-set" of 4D vectors as a result of a union of the standard 4D-FCHC $C_4$ and its rotated realization $C_4^+$:

$$B_4 \equiv C_4 \cup C_4^+ \tag{32}$$

Clearly, the super-lattice set retains the symmetry and the order of isotropy from each of its two constitutive sub-lattices. Specifically, since both $C_4$ and $C_4^+$ are 4$^{th}$ order isotropic, their union is automatically at least 4$^{th}$-order isotropic. Furthermore, direct evaluation of the 6$^{th}$ order moment component in $B_4$ shows that,

$$M_{iiiiii}^{(6)} = 30 \tag{33}$$

Since the requirement (18) for 6$^{th}$ order isotropy is

$$M_{iiiiii}^{(6)} = 15 \frac{bc^6}{D(D+2)(D+4)} = 30 \tag{34}$$

where $b = 48$, $D = 4$, and $c = \sqrt{2}$, we find that the super set $B_4$ is indeed isotropic up to $N = 6$.

Thus, we have described here a systematic procedure to construct higher order isotropic lattice velocity sets. From (11), we see that a $(n+2)^{nd}$ order lattice can be formed out of combination of a group of topologically identical $n^{th}$ order lattice sets, each of which is a rotated realization of the others in the group. Using this procedure, we have explicitly constructed a 6$^{th}$ order isotropic lattice vector set.

## 5. Rescaling of lattice velocities to integer values

As pointed out in Sec. 1, one of the advantages of standard LBM is that advection of a particle distribution per time step lands exactly on a node of a lattice. This means that each Cartesian component of every vector in the lattice velocity set must be an integer (in units of lattice spacing, and the lattice convention, $\Delta x = \Delta t = 1$.) However, the lattice sets constructed above do not have integer component values in general. Although non-integer valued velocities can be handled *via* various numerical interpolation schemes (cf. [8]), algorithmic simplicity, and more importantly the exactness of advection is lost by interpolations. In this Section, we describe a way to recover integer-component lattice velocities by an isotropy-preserving rescaling procedure. This is made possible due to the following basic property:

*Basic Property 1: If a set $B$ is a weighted union of $N^{th}$ order isotropic lattice vector sets $A_1, A_2, ..., A_n$, then its $N^{th}$ order moment is a weighted summation of the $N^{th}$ order moments of its sub-lattices:*

$$M^{(N)}(B) = w(A_1)M^{(N)}(A_1) + w(A_2)M^{(N)}(A_2) + ... + w(A_n)M^{(N)}(A_n)$$

*where the constants $w(A_1), w(A_2), ..., w(A_n)$ are weighting factors.*

It is easily see that the resulting moment for $B$ is at least $N^{th}$ order isotropic for any arbitrary choice of these weighting factor values. On the other hand, we expect it may achieve higher than $N^{th}$ order isotropy with some suitable choice of the weighting factors.

### a. Rescaling the 2D octagonal lattice

For simplicity, we first analyze the 2D octagonal lattice $B_2$ in which four of its diagonal velocities originated from $C_2^+$ (*cf.* (23)) have non-integer component values. The task is to rescale these velocities into integer valued ones while preserving the desired isotropy. It is cleat that all topological and symmetry properties of a vector set are unaffected if all its velocities are rescaled by a common scalar factor. In particular, if all vectors in $C_2^+$ are multiplied by $\sqrt{2}$, we obtain a new lattice set $\tilde{C}_2^+$ having four integer valued velocities,

$$\tilde{C}_2^+ = \{(\pm 1, \pm 1)\} \tag{35}$$

These new vectors simply correspond to the so called "diagonal" links on a square lattice.

We now form an alternative lattice superset $\tilde{B}_2$ out of a union of $C_2$ and the rescaled lattice $\tilde{C}_2^+$. Note that this new superset no longer possesses the $8^{th}$ degree rotational symmetry of the original octagonal lattice. The goal here is to re-establish the $4^{th}$ order moment isotropy for $\tilde{B}_2$ by choosing the proper weight factors ($w_\alpha$) for the two square lattices $C_2$ and $\tilde{C}_2^+$.

For this purpose we observe first that, though topologically identical, an $N^{th}$ order moment tensor in a rescaled set differs by a constant factor from its original pre-rescaled value, and that this constant factor is the $N^{th}$ power of the rescaling factor,

$$\mathbf{M}^{(N)}(\tilde{C}_2^+) = \sqrt{2}^N \mathbf{M}^{(N)}(C_2^+) = 2^{N/2} \mathbf{M}^{(N)}(C_2^+) \tag{36}$$

Therefore, for $4^{th}$ order moment tensors, we have

$$\mathbf{M}^{(4)}(\tilde{C}_2^+) = 4\mathbf{M}^{(4)}(C_2^+) \tag{37}$$

Next, using the Basic Property 1, since $C_2$ and $\tilde{C}_2^+$ are both $2^{nd}$ order isotropic, their union $\tilde{B}_2$ is also at least $2^{nd}$ order isotropic for any arbitrary choice of the weighting factor values. This is explicitly demonstrated by evaluating the $2^{nd}$ order moment,

$$\mathbf{M}^{(2)}(\tilde{B}_2) \equiv \sum_{\alpha=1}^{b} w_\alpha \mathbf{c}_\alpha \mathbf{c}_\alpha = w(C_2)\mathbf{M}^{(2)}(C_2) + w(\tilde{C}_2^+)\mathbf{M}^{(2)}(\tilde{C}_2^+)$$

$$= w(C_2)\mathbf{M}^{(2)}(C_2) + \sqrt{2}^2 w(\tilde{C}_2^+)\mathbf{M}^{(2)}(C_2^+)$$

$$= (w(C_2) + 2w(\tilde{C}_2^+))\mathbf{M}^{(2)}(C_2) \tag{38}$$

The last equality holds because the second-order moments of $C_2$ and $C_2^+$ are equal.

On the other hand, with appropriately chosen weighting factors a superset can be shown to possess higher isotropy than those of its constitutive subsets. Specifically, one can construct a $4^{th}$ order isotropic $\tilde{B}_2$ out of the $2^{nd}$-order isotropic $C_2$ and $\tilde{C}_2^+$ by choosing the weighting factors $w(C_2) = 1$ and $w(\tilde{C}_2^+) = 1/4$. Indeed,

$$\mathbf{M}^{(4)}(\tilde{B}_2) \equiv \sum_{\alpha=1}^{b} w_\alpha \mathbf{c}_\alpha \mathbf{c}_\alpha \mathbf{c}_\alpha \mathbf{c}_\alpha$$

$$= w(C_2)\mathbf{M}^{(4)}(C_2) + w(\tilde{C}_2^+)\mathbf{M}^{(4)}(\tilde{C}_2^+)$$

$$= w(C_2)\mathbf{M}^{(4)}(C_2) + \sqrt{2}^4 w(\tilde{C}_2^+)\mathbf{M}^{(4)}(C_2^+)$$

$$= \mathbf{M}^{(4)}(C_2) + \mathbf{M}^{(4)}(C_2^+) = \mathbf{M}^{(4)}(B_2) \tag{39}$$

where we have used (37). Recall from the previous section that $B_2$ is $4^{th}$ order moment isotropic, so we have shown that $\tilde{B}_2$ is also isotropic up to $4^{th}$ order. It is interesting to point out that lattice set $\tilde{B}_2$ together with its weighting factors corresponds precisely to the discrete velocities (excluding the zero vector) in the so called D2Q9 lattice model [19]. The procedure described here provides a geometrical understanding as to why such a lattice set is 4$^{th}$ order isotropic.

*b. Rescaling of the combined FCHC*

We can also examine possible rescaling of $B_4$ in order to produce an integer valued set. As above, the key step is to properly rescale the non-integer sub-set $C_4^+$ into a new set $\tilde{C}_4^+$ containing only integer-component velocities. This is accomplished by multiplying each velocity in $C_4^+$ by $\sqrt{2}$, so that the rescaled set $\tilde{C}_4^+$ has 24 new integer valued velocity vectors,

$$\tilde{C}_4^+ = \{(\pm 2,0,0,0),(0,\pm 2,0,0),(0,0,\pm 2,0),(0,0,0,\pm 2),$$

$$(\pm 1,\pm 1,\pm 1,\pm 1)\} \tag{40}$$

Once again, the rescaled lattice set has all the same topological structure as $C_4^+$ except the resulting moment at each order differs by a constant factor. For instance,

$$\mathbf{M}^{(6)}(\tilde{C}_4^+) = \sqrt{2}^6 \mathbf{M}^{(6)}(C_4^+) = 8\mathbf{M}^{(6)}(C_4^+) \tag{41}$$

We can obtain a new 6$^{th}$ order isotropic super-lattice $\tilde{B}_4$ by forming a properly weighted union of $C_4$ and $\tilde{C}_4^+$. By carrying out steps similar to those used for the 2D lattice, the proper weights can easily be shown to be $w(C_4) = 1$ and $w(\tilde{C}_4^+) = 1/8$. As a result, the new super-lattice set $\tilde{B}_4$ has 48 ($b = 48$) integer-component velocities when $D = 4$. Projecting this onto a 3-dimensional subspace, we can show (see Appendix 2) that $\tilde{B}_4$ reduces to the so called 34-state LBM model at temperature $T = 2/5$ satisfying 6$^{th}$ order moment isotropy [11,12, 25].

## 5. Mach number expansions for equilibrium distributions

It is well known that the standard equilibrium distribution in LBE usually takes a polynomial form in powers of the fluid velocity, $\mathbf{u}$, that is commonly interpreted as a small Mach number expansion of the Maxwell-Boltzmann equilibrium [12, 22, 27]:

$$f_\alpha^{eq} = w_\alpha \rho \, exp[(\mathbf{c}_\alpha \cdot \mathbf{u})/T] \, exp[-\mathbf{u}^2/2T]$$

$$= w_\alpha \rho \, [1 + \frac{\mathbf{c}_\alpha \cdot \mathbf{u}}{T} + \frac{(\mathbf{c}_\alpha \cdot \mathbf{u})^2}{2T^2} + \frac{(\mathbf{c}_\alpha \cdot \mathbf{u})^3}{6T^3}$$

$$+ \frac{(\mathbf{c}_\alpha \cdot \mathbf{u})^4}{24T^4} + \cdots][1 - \frac{\mathbf{u}^2}{2T} + \frac{\mathbf{u}^4}{8T^2} + \cdots] \tag{42}$$

where $\rho$, $T$ and $\mathbf{u}$ are, respectively, the local fluid density, temperature and velocity. For some typical lattices such as the isothermal LBM models D3Q15 and D3Q19, the temperature $T = 1/3$ [19], while for the 34-state model $T$ can vary between 1/3 and 2/3 [12]. In fact, most LBM models have the above form further restricted to only include terms up to $O(\mathbf{u}^2)$, so that

-14-

$$f_\alpha^{eq} = w_\alpha \rho [1 + \frac{\mathbf{c}_\alpha \cdot \mathbf{u}}{T} + \frac{(\mathbf{c}_\alpha \cdot \mathbf{u})^2}{2T^2} - \frac{\mathbf{u}^2}{2T}] \tag{43}$$

The reason for using this $O(\mathbf{u}^2)$ truncation is partly historical, but it is also because higher power terms require moment isotropy beyond 4$^{th}$ order, which most existing LBM models lack.

Having 6$^{th}$ order isotropy available, we can now show that the expanded form in (42) can retain terms up to $O(\mathbf{u}^5)$ if mass and momentum conservation and the correct non-viscous momentum flux form are satisfied. Retaining terms up to $O(\mathbf{u}^5)$, (42) gives

$$\begin{aligned} f_\alpha^{eq} = w_\alpha \rho [&1 + \frac{\mathbf{c}_\alpha \cdot \mathbf{u}}{T} + \frac{(\mathbf{c}_\alpha \cdot \mathbf{u})^2}{2T^2} - \frac{\mathbf{u}^2}{2T} \\ &+ \frac{(\mathbf{c}_\alpha \cdot \mathbf{u})^3}{6T^3} - \frac{(\mathbf{c}_\alpha \cdot \mathbf{u})\mathbf{u}^2}{2T^2} \\ &+ \frac{(\mathbf{c}_\alpha \cdot \mathbf{u})^4}{24T^4} - \frac{(\mathbf{c}_\alpha \cdot \mathbf{u})^2 \mathbf{u}^2}{4T^3} + \frac{\mathbf{u}^4}{8T^2} \\ &+ \frac{(\mathbf{c}_\alpha \cdot \mathbf{u})^5}{120T^5} - \frac{(\mathbf{c}_\alpha \cdot \mathbf{u})^3 \mathbf{u}^2}{12T^4} + \frac{(\mathbf{c}_\alpha \cdot \mathbf{u})\mathbf{u}^4}{8T^3}] \end{aligned} \tag{44}$$

For the zero speed ($\mathbf{c}_\alpha \equiv 0$) velocity state, (44) leads to the simple positive-definite form:

$$f_0^{eq} = w_0 \rho [1 - \frac{\mathbf{u}^2}{2T} + \frac{\mathbf{u}^4}{8T^2}] > 0$$

This is immediately recognized to be beneficial: Unlike the conventional second-order form, this never becomes negative for any velocity value $\mathbf{u}$.

It is directly verifiable that the higher-order expanded equilibrium form satisfies the requirements for correct fundamental hydrodynamics, if the lattice velocity set meets the following conditions:

$$\sum_{\alpha=1} w_\alpha + w_0 = 1$$

$$\sum_{\alpha=1} w_\alpha c_{\alpha,i} c_{\alpha,j} = T \delta_{ij}$$

$$\sum_{\alpha=1} w_\alpha c_{\alpha,i} c_{\alpha,j} c_{\alpha,k} c_{\alpha,l} = T^2 \Delta_{ijkl}^{(4)}$$

$$\sum_{\alpha=1} w_\alpha c_{\alpha,i} c_{\alpha,j}\ c_{\alpha,k} c_{\alpha,l} c_{\alpha,m} c_{\alpha,n} = T^3 \Delta_{ijklmn}^{(6)} \tag{45}$$

while moments involving any odd powers of $\mathbf{c}_\alpha$ vanish. Combining (44) and (45), one can immediately verify that,

$$\sum_{\alpha=0} f_\alpha^{eq} = \rho$$

$$\sum_{\alpha=1} \mathbf{c}_{\alpha,i} f_\alpha^{eq} = \rho u_i$$

$$\sum_{\alpha=1} \mathbf{c}_\alpha^2 f_\alpha^{eq} = \rho \mathbf{u}^2 + DT$$

$$\sum_{\alpha=1} \mathbf{c}_{\alpha,i} \mathbf{c}_{\alpha,j} f_\alpha^{eq} = \rho u_i u_j + \rho T \delta_{ij} \quad (46)$$

The first three expressions correspond to the correct mass, momentum and energy in a physical fluid, while the last expression is necessary for ensuring the correct hydrodynamic momentum flux up to $O(\mathbf{u}^5)$.

Note that, if (44) only retains terms up to $O(\mathbf{u}^4)$ as opposed to $O(\mathbf{u}^5)$, the correct energy flux tensor form is retained:

$$\sum_{\alpha=1} \mathbf{c}_{\alpha,i} \mathbf{c}_{\alpha,j} \mathbf{c}_{\alpha,k} f_\alpha^{eq} = \rho u_i u_j u_k + \rho T [u_i \delta_{jk} + u_j \delta_{ki} + u_k \delta_{ij}] \quad (47)$$

On the other hand, commonly used LBM models such as those labeled as "DmQn" (see Appendix 4) satisfy the relation in (45) only up to $T^2$, so that they do not give the physically correct energy flux form for thermo-hydrodynamics.

The conditions (45) indicate that, not only are isotropy requirements necessary, moments using discrete velocities must also satisfy some specific scalar values in powers of temperature $T$ [12]. Based on the above analysis, we expect (without proof) the existence of a general relationship at all orders between the power of the Mach number expansion for the equilibrium distribution and the order of lattice velocity set: *The expansion of the equilibrium distribution can be carried out to $O(\mathbf{u}^{2M-1})$, if the isotropy conditions (45) are expanded to $2M^{th}$ order:*

$$\sum_{\alpha=1} w_\alpha c_{\alpha,i_1} \cdots c_{\alpha,i_{2m}} = T^m \Delta_{i_1 \cdots i_{2m}}^{(2m)}, \quad m = 1, \ldots, M \quad (48)$$

*where $M$ is a postive integer.* Indeed, this has been specifically verified up to 8[th] order, and is expected to hold for any order.

This general property can also be expressed in terms of two separate sets of conditions: One governs the lattice isotropy at each speed level, and the other defines a set of relationships among various speed levels. Explicitly, for a lattice obeying isotropy up to order $2M$ at each speed level, there exists a set of conditions determined by (10). This together with (48) immediately indicates an additional set of scalar constraint relationships among different speed levels up to order $2M$, see [12]:

$$\sum_{\alpha=0} \mathbf{c}_\alpha^{2m} w_\alpha = D(D+2) \cdots (D+2(m-1)) T^m, \quad m = 1, 2, \ldots, M \quad (49)$$

The general constraint relations of (48) [or, equivalently, (10) and (49)] automatically guarantee correct hydrodynamic moments (i.e., the same as that of continuous Boltzmann kinetic theory) up to order $2M$-1. There is no need for *a posteriori* coefficient matching (as in early-day LBM theoretical model formulations).

*a. Sixth order LBM via multi-speed levels*

We present in this subsection a construction of a particular velocity set that satisfies the requirements in (45) up to sixth order. This is accomplished by using three different speed levels, each of which is a simple lattice obeying 6[th] order isotropy as in (10). That is, for each speed level $\beta$

$$M^{(n),\beta}_{i_1 i_2 \ldots i_n} = \frac{c^n_\beta b(D-2)!!}{(D+n-2)!!} \Delta^{(n)}_{i_1 i_2 \ldots i_n}; \quad n = 2, 4, 6 \tag{50}$$

For example, using the methods discussed in the preceding sections, one can verify that any of the following three simple $D = 4$ lattice velocity sets meet such a requirement. They are

**Set $\beta = 1$:**

$$B_4^{\beta=1} = \{(\pm 1, \pm 1, 0, 0), (\pm 1, 0, \pm 1, 0), (\pm 1, 0, 0, \pm 1),$$

$$(0, \pm 1, \pm 1, 0), (0, \pm 1, 0, \pm 1), (0, 0, \pm 1, \pm 1),$$

$$(\pm\sqrt{2}, 0, 0, 0), (0, \pm\sqrt{2}, 0, 0), (0, 0, \pm\sqrt{2}, 0),$$

$$(0, 0, 0, \pm\sqrt{2}), (\pm\sqrt{1/2}, \pm\sqrt{1/2}, \pm\sqrt{1/2}, \pm\sqrt{1/2})\} \tag{51}$$

**Set $\beta = 2$:**

$$B_4^{\beta=2} = \{(\pm 2, 0, 0, 0), (0, \pm 2, 0, 0), (0, 0, \pm 2, 0), (0, 0, 0, \pm 2),$$

$$(\pm 1, \pm 1, \pm 1, \pm 1),$$

$$(\pm\sqrt{2}, \pm\sqrt{2}, 0, 0), (\pm\sqrt{2}, 0, \pm\sqrt{2}, 0),$$

$$(\pm\sqrt{2}, 0, 0, \pm\sqrt{2}), (0, \pm\sqrt{2}, \pm\sqrt{2}, 0),$$

$$(0, \pm\sqrt{2}, 0, \pm\sqrt{2}), (0, 0, \pm\sqrt{2}, \pm\sqrt{2})\} \tag{52}$$

**Set $\beta = 3$:**

$$B_4^{\beta=3} = \{(\pm 2, \pm 2, 0, 0), (\pm 2, 0, \pm 2, 0), (\pm 2, 0, 0, \pm 2),$$

$$(0, \pm 2, \pm 2, 0), (0, \pm 2, 0, \pm 2), (0, 0, \pm 2, \pm 2),$$

$$(\pm 2\sqrt{2}, 0, 0, 0), (0, \pm 2\sqrt{2}, 0, 0), (0, 0, \pm 2\sqrt{2}, 0), (0, 0, 0, \pm 2\sqrt{2}),$$

$$(\pm\sqrt{2}, \pm\sqrt{2}, \pm\sqrt{2}, \pm\sqrt{2})\} \tag{53}$$

Each of these three simple lattice sets contains 48 distinct vector values. That is, $b_\beta = 48$, $\beta = 1, 2, 3$. Notice also that all velocity vectors within each of the three lattice sets have the same magnitude,

$$\mathbf{c}^2_{\alpha,\beta} = c^2_\beta, \quad \forall \alpha = 1, 2, \ldots, b_\beta, \quad \beta = 1, 2, 3$$

although they are different for different speed levels. Specifically, $c^2_{\beta=1} = 2$, $c^2_{\beta=2} = 4$ and $c^2_{\beta=3} = 8$. Therefore, the weights within each of the three lattice sets are the same but different from one set to another:

$$w_{\alpha,\beta} = w_\beta, \quad \forall \alpha = 1,2,...,b, \ \beta = 1,2,3$$

Using the rescaling procedure described above, we can construct three new lattice sets based on the current ones, and they are,

**Set $\beta = 1$:**

$$\tilde{B}_4^{\beta=1} = \{(\pm 1, \pm 1, 0, 0), (\pm 1, 0, \pm 1, 0), (\pm 1, 0, 0, \pm 1),$$
$$(0, \pm 1, \pm 1, 0), (0, \pm 1, 0, \pm 1), (0, 0, \pm 1, \pm 1),$$
$$(\pm 2, 0, 0, 0), (0, \pm 2, 0, 0), (0, 0, \pm 2, 0), (0, 0, 0, \pm 2),$$
$$(\pm 1, \pm 1, \pm 1, \pm 1)\} \tag{54}$$

**Set $\beta = 2$:**

$$\tilde{B}_4^{\beta=2} = \{(\pm 2, 0, 0, 0), (0, \pm 2, 0, 0), (0, 0, \pm 2, 0), (0, 0, 0, \pm 2),$$
$$(\pm 1, \pm 1, \pm 1, \pm 1),$$
$$(\pm 2, \pm 2, 0, 0), (\pm 2, 0, \pm 2, 0), (\pm 2, 0, 0, \pm 2),$$
$$(0, \pm 2, \pm 2, 0), (0, \pm 2, 0, \pm 2), (0, 0, \pm 2, \pm 2)\} \tag{55}$$

**Set $\beta = 3$:**

$$\tilde{B}_4^{\beta=3} = \{(\pm 2, \pm 2, 0, 0), (\pm 2, 0, \pm 2, 0), (\pm 2, 0, 0, \pm 2),$$
$$(0, \pm 2, \pm 2, 0), (0, \pm 2, 0, \pm 2), (0, 0, \pm 2, \pm 2),$$
$$(\pm 4, 0, 0, 0), (0, \pm 4, 0, 0), (0, 0, \pm 4, 0), (0, 0, 0, \pm 4),$$
$$(\pm 2, \pm 2, \pm 2, \pm 2)\} \tag{56}$$

so that all velocity vectors in all these rescaled sets are integer-valued. It is easily recognized that the new lattice sets are realized by multiplying the non-integer valued vectors in the original sets by $\sqrt{2}$. According to the rescaling procedure shown previously for preserving the 4$^{th}$ order isotropy, the only step required is to have the corresponding weights for these rescaled states to be changed to $w'_\beta = w_\beta/\sqrt{2}^6 = w_\beta/8$.

Together with an additional zero velocity state (i.e., (0,0,0,0)), we can determine the values for these weights by enforcing the fundamental constraints (49). Specifically, using $D = 4$ and $b = 48$, we have

$$w_0 + 24[(1+\frac{1}{8})w_1 + (1+\frac{1}{8})w_2 + (1+\frac{1}{8})w_3] = 1$$

$$15(w_1 + 2w_2 + 4w_3) = T$$

$$6(w_1 + 2^2 w_2 + 4^2 w_3) = T^2$$
$$2(w_1 + 2^3 w_2 + 4^3 w_3) = T^3 \tag{57}$$

where $w_0$ is the weighting factor for the zero velocity state, that is necessary for satisfying the first constraint above. Using straightforward algebra, we obtain

$$w_0 = 1 - 27(w_1 + w_2 + w_3)$$

$$w_1 = T[\frac{8}{45} - \frac{T}{3} + \frac{T^2}{6}]$$

$$w_2 = \frac{T}{24}[-\frac{8}{5} + 5T - 3T^2]$$

$$w_3 = \frac{T}{48}[\frac{4}{15} - T + T^2] \tag{58}$$

There exist solutions having positive real values for all these weights within the range of temperature, $0.44 < T < 1.23$.

The above form a specific lattice set that is both 6$^{th}$ order isotropic and satisfies the fundamental hydrodynamic constraints (45). It consists of 97 ($=1+4\times 24$) distinct 4-dimensional vectors. Fortunately, in real applications only $D=3$ or lower is required. Projecting this set onto 3-dimensional space results in fewer distinct velocity values due to degeneracy. The explicit expression for the projected lattice velocity set $B_3$ in 3D is given as the union of the following six level sub-sets together with their associated weights, as listed below:

One zero velocity "(000)" state, $\mathbf{c}_0 = (0,0,0)$ and its associated weighting factor

$$w_{000} = w_0 + \frac{1}{4}w_1 + 2w_2 + \frac{1}{4}w_3;$$

Twelve "(110)" states,

$$\{(\pm 1, \pm 1, 0), (\pm 1, 0, \pm 1), (0, \pm 1, \pm 1)\}$$

all having the weighting factor $w_{110} = w_1$;

Six "(100)" states,

$$\{(\pm 1, 0, 0), (0, \pm 1, 0), (0, 0, \pm 1)\}$$

with $w_{100} = 2w_1$;

Eight "(111)" states

$$\{(\pm 1, \pm 1, \pm 1)\}$$

with $w_{111} = \frac{1}{4}w_1 + 2w_2$;

Six "(200)" states

$$\{(\pm 2, 0, 0)\}$$

with

$$w_{200} = \frac{1}{8} w_1 + \frac{5}{4} w_2 + 2 w_3;$$

Twelve "(220)" states

$$\{(\pm 2, \pm 2, 0), (\pm 2, 0, \pm 2), (0, \pm 2, \pm 2)\}$$

with $w_{220} = \frac{1}{8} w_2 + w_3$;

Eight "(222)" states

$$\{(\pm 2, \pm 2, \pm 2)\}$$

with $w_{222} = \frac{1}{4} w_3$;

And six "(400)" states

$$\{(\pm 4, 0, 0), (0, \pm 4, 0), (0, 0, \pm 4)\}$$

with $w_{400} = \frac{1}{8} w_3$. This gives a total of 59 distinct velocity states in the 3D projected set and offers a specific LBM model for achieving $6^{th}$ order moment requirements in (45) for either thermohydrodynamics or isothermal Burnett effects.

## 6. Discussion

In this paper, we have presented a systematic analysis of the relationships between the degree of rotational symmetry of a given discrete vector set and the order of isotropy in its resulting moment tensors. The latter constitutes the core ingredient in formulating lattice Boltzmann models. For this purpose, we have derived a set of necessary and sufficient conditions for measuring the order of isotropy for any given lattice vector set. Using these, we are able to construct higher-order isotropic discrete velocity sets necessary to capture physical effects beyond that of isothermal Navier-Stokes fluids. This removes a key limitation on conventional lattice Boltzmann models that only satisfy $4^{th}$ order moment isotropy, and thus only produce correct momentum flux tensors up to those required by the isothermal Navier-Stokes physics. Both Navier-Stokes thermal energy fluxes or higher-order non-equilibrium moment fluxes require isotropy to at least $6^{th}$ order [10].

The present work also provides direct geometric insights into some popularly used lattice models, such as the well known D3Q15 and D3Q19. This geometric understanding allows us to construct higher-order symmetric lattice Boltzmann models systematically. Furthermore, we have provided a description of an isotropy preserving scheme that rescales the magnitude of each velocity in a lattice set, so that the resulting vectors are all have integer components. This is desirable for lattice Boltzmann models in order to perform the advection step in a simple and accurate manner. Finally, we have provided a set of sufficient conditions for achieving correct physical hydrodynamics at any moment order. Satisfying these can automatically guarantee the right physical flow behavior as opposed to relying on *a posteriori* coefficient matching to recover known hydrodynamic equations. The latter is in general not available beyond the isothermal Navier-Stokes level.

It is important to mention the alternative systematic theoretical formulation developed by Shan *et al* [21, 22], which is based on an expansion procedure with the Hermite polynomials which form an orthornormal basis in Hilbert space in terms of the hydrodynamic moments. The order of accuracy is explicitly related to the level of truncation for the infinite series. Its associated quadrature at each given truncated level defines a set of discrete points in velocity space resulting in a set of the discrete velocity vectors for LBM. Usually the discrete velocity set generated from such an alternative Hermite expansion and Gaussian quadrature based approach has fewer points. On the other hand, its velocity values are often non-integers. Nevertheless, using a special Cartesian quadrature scheme [21], integer valued discrete velocity values can be constructed [22].

Finally, the proof for the rotational symmetry and moment isotropy relationships in Appendix 1 is only valid for 2-dimensional situations involving velocities of the same magnitude. It is known that such a crystallographic geometric structure is associated with the so-called Platonic solids. For 2D, there are an infinite number of distinct shapes of this type. Unfortunately, there are only a finite number of shapes in dimensions higher than 2. In other words, there does not appear to exist any platonian solid possessing infinite degrees of symmetry in higher dimensional spaces. Therefore, the proof presented here cannot be straightforwardly extended to dimensions higher than two. On the other hand, as shown in this paper, there may exist non-platonian solids involving multiple velocity magnitude values. These may be realized *via* combinations of lower order platonian ones with different velocity magnitude, or projections from higher dimensional spaces. Theoretical studies in this regard, now planned for the future, have potential implications both in fundamental physics as well as in the formulation of higher order discrete Boltzmann models.

**Acknowledgments**: We are grateful to Raoyang Zhang, Xiaowen Shan, and Ilya Staroselsky for important discussions. This work is supported in part by National Science Foundation.

# Appendix 1: Degree of rotational symmetry and order of moment isotropy in two dimensions

Here we consider a lattice velocity set, $C$, containing $b$ discrete velocity values in 2D, in which all velocities have the same velocity magnitude $c$. The set forms a crystallographic structure that is $b$-order rotationally symmetric. That is, when it undergoes a rigid rotation of an angle which is a multiple of $2\pi/b$, the rotated set returns to the original set. In addition, we still assume that the lattice set is parity invariant. That is, it is invariant if every velocity in the set is reversed, $\mathbf{c}_\alpha \to -\mathbf{c}_\alpha$, $\alpha = 1,...,b$.

Following the main text, an $n^{\text{th}}$ order basis moment tensor is given as,

$$\mathbf{M}^{(n)} \equiv \sum_\alpha^b \underbrace{\mathbf{c}_\alpha \mathbf{c}_\alpha \cdots \mathbf{c}_\alpha}_{n} \tag{59}$$

Taking repeated scalar products between this tensor and a vector $\mathbf{v}$, we have

$$\mathbf{M}^{(n)} \cdot \mathbf{v} = \sum_\alpha^b \sum_{i_1=x}^y \sum_{i_2=x}^y \cdots \sum_{i_n=x}^y c_{\alpha,i_1} c_{\alpha,i_2} \cdots c_{\alpha,i_n} v_{i_1} v_{i_2} \cdots v_{i_n}$$

$$= \sum_{\alpha}^{b} (\mathbf{c}_{\alpha} \cdot \mathbf{v})^n \tag{60}$$

It is easily recognized that $\mathbf{M}^{(n)}$ is isotropic if and only if for any arbitrary 2D vector $\mathbf{v}$, the following is true,

$$\sum_{\alpha}^{b} (\mathbf{c}_{\alpha} \cdot \mathbf{v})^n \propto v^n$$

where $v = |\mathbf{v}|$ is the magnitude of $\mathbf{v}$. In other words, defining an unity vector, $\hat{v} = \mathbf{v}/|\mathbf{v}|$, the above condition is equivalent to,

$$\sum_{\alpha}^{b} (\mathbf{c}_{\alpha} \cdot \hat{v})^n = A \tag{61}$$

where $A$ must be a constant. Here $\hat{v} = \mathbf{v}/|\mathbf{v}|$.

It is easily observed that a $b$-fold rotationally symmetric 2D vector set can be, without loss of generality, expressed as

$$C = \{\mathbf{c}_{\alpha} = (cos(\frac{2\pi\alpha}{b}), sin(\frac{2\pi\alpha}{b}); \quad \alpha = 0,...,b-1\} \tag{62}$$

To satisfy parity invariance, $b$ must be even. An arbitrary 2D unity vector can also be expressible as, $\hat{v} = (cos(\theta), sin(\theta))$. The isotropy condition in (61) is thus equivalent to having $h_b^{(n)}(\theta)$ (defined below) to be independent of $\theta$.

$$h_b^{(n)}(\theta) \equiv \sum_{\alpha=0}^{b-1} cos^n(\frac{2\pi\alpha}{b} - \theta) \tag{63}$$

It is convenient to rewrite $h_b^{(n)}$ as follows,

$$h_b^{(n)}(\theta) = \frac{1}{2^n} \sum_{\alpha=0}^{b-1} \sum_{j=0}^{n} (\frac{n!}{j!(n-j)!} e^{i\frac{2\pi\alpha}{b}(2j-n)} e^{-i\theta(2j-n)} \tag{64}$$

Since terms that have $j = n/2$ are automatically $\theta$-independent, the condition that (64) is independent of $\theta$ is that for $j \neq n/2$

$$\sum_{\alpha=0}^{b-1} e^{i\frac{2\pi\alpha}{b}(2j-n)} = 0 \tag{65}$$

Equivalently, $\frac{(2j-n)}{b}$ cannot be a non-zero integer for $j = 0,1,...,n$. Since $n$ is even, let $J \equiv n/2$, with $J$ an integer. Therefore, the condition is that $2\frac{j-J}{b}$ is not a non-zero integer for $j = 0,1,...,2J$. Clearly, this is satisfied when $J < b/2$ or $n < b$.

Hence we have shown the following theorem:

*For a 2D parity invariant lattice velocity set that has a $b$-fold rotational symmetry, its moments $\mathbf{M}^{(n)}$ are isotropic for $n < b$.*

In summary, we see that the 4-fold rotational symmetric square lattice only gives moment isotropy to $2^{nd}$-order, the 6-fold hexagonal lattice achieves moment isotropy to $4^{th}$ order, and the 8-fold octagonal lattice has isotropy to 6$^{th}$ order.

## Appendix 2: Isotropy preserving projection of lattice velocities to lower dimensions

It is well known that lower dimensional lattices can be realized *via* projection of higher dimensional ones onto lower dimensional sub-spaces. For example, the velocities in the popular D3Q19 LBE can be viewed as a result of a simple projection of the standard 4D-FCHC (i.e., $C_4$) onto three dimensions. This is accomplished by taking only the first three Cartesian components from the 4D vectors in FCHC. It is easily realized that there could be a multiple of 4D vectors that have the same first three Cartesian component values. Due to such degeneracy, there are only 18 distinct 3D vectors in the resulting projected set instead of 24 in the original set in 4D space. Such a many-to-one mapping defines a degeneracy factor associated with each resulting 3D vector. This factor serves as the weighting factor $w_\alpha$ in the moments calculations (see (7)). This, for instance, explains why the $w_\alpha$ value for lattice velocity $(1,0,0)$ is twice that of $(1,1,0)$ in D3Q19, since the former corresponds to two lattice velocities (i.e., $\{(1,0,0,\pm 1)\}$) in $C_4$. Following this line of argument, one can further view D2Q9 as a projection of D3Q19 onto 2-dimensions. It is straightforward to recognize that a projection procedure defined in this way preserves all the moment properties (both in magnitude and in tensor structure) of the original pre-projected higher dimensional lattices. For instance, all the $w_\alpha$ weighted moments in D3Q19 are exactly the same as those for $C_4$. Consequently, since 4D-FCHC is 4$^{th}$ order isotropic, D3Q19 (as well as D2Q9) is also 4$^{th}$ order isotropic.

In this way, we can also form a clear geometrical understanding of the so called D3Q15 lattice and its order of isotropy. In fact, we can show D3Q15 is a result of projecting the scaled-rotated 4D-FCHC (i.e., $\tilde{C}_4^+$) together with the rescaling procedure defined in the main text of this paper. Directly projecting $\tilde{C}_4^+$ onto 3D gives the following set $Q$ of 15-velocities,

$$\{(\pm 2,0,0),(0,\pm 2,0),(0,0,\pm 2),(0,0,0),(\pm 1,\pm 1,\pm 1)\}$$

in which there are six "speed 2" velocities,

$$\{(\pm 2,0,0),(0,\pm 2,0),(0,0,\pm 2)\}$$

and eight "tri-diagonal" velocities,

$$\{(\pm 1,\pm 1,\pm 1)\}$$

plus a "zero-speed" velocity state $(0,0,0)$. If denoting the set of speed-2 velocities as $T$ and the set of tri-diagonal velocities as $U$, then we can express $Q = T \cup U$. Due to degeneracy of the projection onto 3D, as explained above $w_\alpha(T) = 1$ for the speed-2 velocities, while $w_\alpha(U) = 2$ for the tri-diagonal ones. Apparently each of the two sub-sets is 2$^{nd}$ order isotropic, so that any weighted union of the two is also guaranteed to be isotropic at least up to 2$^{nd}$ order. However, since $Q$ is a direct projection of $\tilde{C}_4^+$, the specific weighted union of $T$ and $U$ in fact has a 4$^{th}$ order moment isotropy.

It is desirable to rescale the speed in $T$ by $1/2$ to form a new set $\tilde{T}$ having speed-1 velocities:

$$\tilde{T} = \{(\pm 1, 0, 0), (0, \pm 1, 0), (0, 0, \pm 1)\}$$

A new set of lattice vectors $\tilde{Q}$ can be then created as a union of $\tilde{T}$ and $U$. However, we must ensure that this new lattice set $\tilde{Q}$ retains $4^{th}$ order isotropy as for $Q$. As explained in the main text of the paper, all the tensor properties of $T$ and $\tilde{T}$ are identical other than some constant scalar factors. Specifically, their $4^{th}$ order moment tensors are related as

$$\mathbf{M}^{(4)}(\tilde{T}) = (\frac{1}{2})^4 \mathbf{M}^{(4)}(T) = \frac{1}{16} \mathbf{M}^{(4)}(T) \tag{66}$$

Therefore, the new lattice vector set $\tilde{Q}$ has the same $4^{th}$ order moment form as that of $Q$ if all the weights in $\tilde{T}$ are related to those in $T$ by

$$w_\alpha(\tilde{T}) = 16 w_\alpha(T)$$

Thus, we must choose

$$w_\alpha(\tilde{T}) = 8 w_\alpha(U)$$

so that $\tilde{Q}$ remains $4^{th}$ order isotropic.

We can recognize that set $\tilde{Q}$ with its above-defined weights precisely corresponds to the non-zero velocities in D3Q15. The analysis above has thus provided geometric understanding as to why D3Q15 is $4^{th}$ order isotropic.

# Appendix 3: Isotropic forms of $2^{nd}$ and $4^{th}$ order moment tensors

An analysis of this problem was given previously by Chris Teixeira (*cf.* [24]). First of all, we establish the fundamental form for an isotropic $2^{nd}$-order moment tensor in $D$-dimensions.:

*Let $\mathbf{v}$ be a random vector in a $D$-dimensional space. If its possible orientations are distributed equally probably (i.e., isotropic) in all directions, then its associated $2^{nd}$-order moment tensor has the following form:*

$$\langle v_\alpha v_\beta \rangle = \frac{\langle v^2 \rangle}{D} \delta_{\alpha\beta} \tag{67}$$

*where $v_\alpha$ and $v_\beta$ are the $\alpha$ and $\beta$ Cartesian components of the $D$-dimensional vector $\mathbf{v}$, respectively. $\delta_{\alpha\beta}$ is a standard Kronecker delta function. In the above, $\langle \bullet \rangle$ represents an average over all possible orientations, and $\langle v^2 \rangle \equiv \langle \mathbf{v}^2 \rangle$.*

**Proof**: By definition,

$$\langle v^2 \rangle = \sum_{\alpha=1}^{D} \langle v_\alpha^2 \rangle \tag{68}$$

Being equally distributed in all directions,

$$\langle v_{\alpha_1}^2 \rangle = \langle v_{\alpha_2}^2 \rangle = \ldots = \langle v_{\alpha_D}^2 \rangle$$

where $\alpha_1, \alpha_2, ..., \alpha_D$ are any Cartesian component indices in $D$-dimensional space. Hence, from (68) we have

$$\langle v^2 \rangle = D \langle v_\alpha^2 \rangle \quad \alpha = 1, 2, ..., D$$

Furthermore, it is easily shown that $\langle v_\alpha v_\beta \rangle = 0$, for any $\alpha \neq \beta$. Combining these results we obtain the form in (67).

Next, we obtain the fundamental form for an isotropic 4$^{th}$-order moment tensor:

*The 4$^{th}$-order moment associated with an isotropically distributed $D$-dimensional vector field* **v** *has the following form*

$$\langle v_\alpha v_\beta v_\gamma v_\delta \rangle = \frac{\langle v^4 \rangle}{D(D+2)} [\delta_{\alpha\beta}\delta_{\gamma\delta} + \delta_{\alpha\gamma}\delta_{\beta\delta} + \delta_{\alpha\delta}\delta_{\gamma\beta}] \tag{69}$$

*where $\alpha$, $\beta$, $\gamma$ and $\delta$ are the Cartesian component indices.* $\langle v^4 \rangle \equiv \langle (\mathbf{v}^2)^2 \rangle$.

**Proof**: It is obvious that $\langle v_\alpha v_\beta v_\gamma v_\delta \rangle$ vanishes unless indices are either all equal or form two distinct pairs. Hence, there are only two possible non-vanishing combinations of indices,

$$\langle v_\alpha^2 v_\beta^2 \rangle \geq 0 \quad \langle v_\alpha^4 \rangle \geq 0 \quad \forall \alpha \neq \beta$$

Based on this consideration, we immediately arrive at the following generic form,

$$\langle v_\alpha v_\beta v_\gamma v_\delta \rangle = A\delta_{\alpha\beta\gamma\delta} + B[\delta_{\alpha\beta}\delta_{\gamma\delta} + \delta_{\alpha\gamma}\delta_{\beta\delta} + \delta_{\alpha\delta}\delta_{\gamma\beta}] \tag{70}$$

where $A$ and $B$ are scalar constants to be determined. Next, straightforward angular integration by parts for an isotropic vector field give

$$\langle v_\alpha^4 \rangle = 3 \langle v_\alpha^2 v_\beta^2 \rangle, \quad \forall \alpha \neq \beta \tag{71}$$

This implies that $A = 0$.

Being isotropically distributed, all components are equivalent so

$$\langle v_\alpha^4 \rangle = \langle v_\beta^4 \rangle, \quad \forall \alpha, \beta = 1, 2, ..., D$$

and

$$\langle v_\alpha^2 v_\beta^2 \rangle = \langle v_\gamma^2 v_\delta^2 \rangle, \quad \alpha \neq \beta, \gamma \neq \delta, \quad \forall \alpha, \beta, \gamma, \delta = 1, 2, ..., D$$

Furthermore,

$$\mathbf{v}^4 \equiv (\sum_{\alpha=1}^{D} v_\alpha^2)^2 = \sum_{\alpha=1}^{D} v_\alpha^4 + 2\sum_{\alpha > \beta} v_\alpha^2 v_\beta^2$$

so averaging over all directions gives

$$\langle v^4 \rangle = \left\langle \sum_{\alpha=1}^{D} v_\alpha^4 \right\rangle + 2\left\langle \sum_{\alpha > \beta} v_\alpha^2 v_\beta^2 \right\rangle$$

$$= D\langle v_\alpha^4 \rangle + D(D-1)\langle v_\alpha^2 v_\beta^2 \rangle, \quad \forall \alpha \neq \beta \tag{72}$$

Using (71), we obtain

$$\langle v^4 \rangle = (3D + D(D-1))\langle v_\alpha^2 v_\beta^2 \rangle, \quad \forall \alpha \neq \beta$$

Simplifying this expression gives

$$\langle v_\alpha^2 v_\beta^2 \rangle = \frac{\langle v^4 \rangle}{D(D+2)}$$

Thus, $B = \langle v^4 \rangle / [D(D+2)]$, which completes the derivation of (69).

## Appendix 4: Moment properties for D3Q19 and D3Q15

We have shown previously that D3Q19 is a projection of the 4D-FCHC plus a zero-speed state. Hence the 2$^{nd}$ and 4$^{th}$ order basic moment tensors are given by the vectors in FCHC:

$$\mathbf{M}^{(2)}(D3Q19)_{ij} = \sum_\alpha w_\alpha c_{\alpha,i} c_{\alpha,j}$$

$$= w^{(1)} M^{(2)}(4DFCHC)_{ij}$$

$$= w^{(1)} \frac{c^2 b}{D} \delta_{ij}$$

$$\mathbf{M}^{(4)}(D3Q19)_{ijkl} = \sum_\alpha w_\alpha c_{\alpha,i} c_{\alpha,j} c_{\alpha,k} c_{\alpha,l}$$

$$= w^{(1)} M^{(4)}(4DFCHC)_{ijkl}$$

$$= w^{(1)} \frac{c^4 b}{D(D+2)} \Delta_{ijkl}^{(4)} \tag{73}$$

where $w^{(1)}$ is the weighting factor (same for all non-zero velocities) for 4D-FCHC lattice vectors. Indices $i$, $j$, $k$ and $l$ correspond to Cartesian components in 3-dimensions.

Using $c^2 = 2$, $b = 24$ and $D = 4$ for lattice vectors in 4D-FCHC, we obtain

$$\mathbf{M}^{(2)}(D3Q19)_{ij} = 12 w^{(1)} \delta_{ij}$$

$$\mathbf{M}^{(4)}(D3Q19)_{ijkl} = 4 w^{(1)} \Delta_{ijkl}^{(4)} \tag{74}$$

Hence in order to satisfy the fundamental hydrodynamic requirement defined in (48), we must choose $w^{(1)}$ so that

$$12 w^{(1)} = T_0, \quad 4 w^{(1)} = T_0^2 \tag{75}$$

for some constant $T_0$. The solution is $T_0 = 1/3$, and $w^{(1)} = 1/36$. Furthermore $M^{(0)}$ is necessary for mass conservation, so

$$w^{(0)} + 24 w^{(1)} = 1 \rightarrow w^{(0)} = 1/3$$

where $w^{(0)}$ is the weighting factor for the zero-speed state. Together with the factor 2 degeneracy for the non-diagonal vectors, this completely defines D3Q19. We have also seen it satisfies moment requirements only up to 4$^{th}$ order.

Next, we look at the moment properties in D3Q15. According to the analysis given in the previous sections (see Appendix 2), we know that $M^{(4)}$ for D3Q15 is equal to that in the scaled 4D-FCHC $\tilde{C}_4^+$. That is,

$$M^{(4)}(D3Q15)_{ijkl} = w^{(1)} M^{(4)}(\tilde{C}_4^+)_{ijkl} = w^{(1)} \frac{c^4 b}{D(D+2)} \Delta^{(4)}_{ijkl}$$

$$= 16 w^{(1)} \Delta^{(4)}_{ijkl} \qquad (76)$$

where $c^2 = 4$, $b = 24$, and $D = 4$. Based on the construction in Appendix 2, the 2$^{nd}$ order basic moment tensor is given by,

$$M^{(2)}(D3Q15)_{ij} = w^{(1)}[16 M^{(2)}(\tilde{T})_{ij} + 2 M^{(2)}(U)_{ij}] \qquad (77)$$

Directly evaluating $M^{(2)}(\tilde{T})_{ij}$ and $M^{(2)}(U)_{ij}$, we get

$$M^{(2)}(\tilde{T})_{ij} = 2\delta_{ij}$$

and

$$M^{(2)}(U)_{ij} = 8\delta_{ij}$$

Thus

$$M^{(2)}(D3Q15)_{ij} = w^{(1)}[32\delta_{ij} + 16\delta_{ij}] = 48 w^{(1)} \delta_{ij} \qquad (78)$$

The solution for $16 w^{(1)} = T_0$ and $48 w^{(1)} = T_0^2$ is $T_0 = 1/3$ and $w^{(1)} = 1/72$. Taking into account the degeneracy and rescaling factors, the weighting factors for vectors in the set $\tilde{T}$ are $w_\alpha(\tilde{T}) = 8/72 = 1/9$; while the weighting factors for vectors in set $U$ are $w_\alpha(U) = 1/72$.

The weighting factor for the zero-speed state is obtained once again *via*

$$w^{(0)} + 6 w_\alpha(\tilde{T}) + 8 w_\alpha(U) = 1$$

which gives $w^{(0)} = 2/9$. All these factor values completely define the popular D3Q15 LBM model. Once again we know from the above analysis that D3Q15 is also only valid for hydrodynamic moments up to 4$^{th}$ order.